\documentclass[aps,prl,reprint,showpacs,showkeys,superscriptaddress,floatfix,twocolumn]{revtex4-2}
\usepackage{multirow}
\usepackage{graphicx}
\usepackage{bm}
\usepackage{dcolumn}
\usepackage[colorlinks=true,linkcolor=blue,urlcolor=blue,citecolor=blue]{hyperref}
\usepackage{natbib}
\usepackage{amsmath}
\usepackage{times}
\usepackage{xcolor}
\usepackage{soul}
\usepackage[normalem]{ulem}

\newcommand{\bybo}{Ba$_3$Yb(BO$_3$)$_3$}
\newcommand{\blbo}{Ba$_3$Lu(BO$_3$)$_3$}

\begin{document}

\title{Realization of Quantum Dipoles in Triangular Lattice Crystal Ba$_3$Yb(BO$_3$)$_3$}

\author{Rabindranath Bag}\affiliation{Department of Physics, Duke University, Durham, NC, USA}
\author{Matthew Ennis}\thanks{equal contribution}\affiliation{Department of Physics, Duke University, Durham, NC, USA}
\thanks{equal contribution}
\author{Chunxiao Liu}\thanks{equal contribution}\affiliation{Department of Physics, University of California, Santa Barbara, CA, USA}
\author{Sachith E. Dissanayake}\affiliation{Department of Physics, Duke University, Durham, NC, USA}
\author{Zhenzhong Shi}\affiliation{Department of Physics, Duke University, Durham, NC, USA}
\author{Jue Liu}\affiliation{Neutron Scattering Division, Oak Ridge National Laboratory, Oak Ridge, TN, USA}
\author{Leon Balents}\affiliation{Kavli Institute for Theoretical Physics, University of California, Santa Barbara, CA, USA}
\author{Sara Haravifard}
\email[email:]{sara.haravifard@duke.edu} \affiliation{Department of Physics, Duke University, Durham, NC, USA} \affiliation{Department of Mechanical Engineering and Materials Science, Duke University, Durham, NC, USA}

\date{\today}

\begin{abstract}
We investigate the thermodynamic properties of the ytterbium-based triangular lattice compound Ba$_3$Yb(BO$_3$)$_3$. The results demonstrate the absence of any long-range ordering down to 56 mK. Analysis of the magnetization, susceptibility and specific heat measurements suggests that Ba$_3$Yb(BO$_3$)$_3$ may realize a S = $\frac{1}{2}$ quantum dipole lattice, in which the dominant interaction is the long range dipole-dipole coupling on the geometrically frustrated triangular lattice, and exchange interactions are subdominant or negligible.
	
\end{abstract}
	
\maketitle

Ytterbium-based layered triangular lattice materials received a significant amount of attention due to presence of an effective spin $\frac{1}{2}$ ground state and strong spin-orbit coupling. The compound YbMgGaO$_4$ in which Yb$^{3+}$ ions form a triangular lattice has been considered to host quantum spin liquid (QSL) ground state \cite{LiPRL2015,LiSciR2015,LiPRB2016,LiPRL2017}. However, debate arises due to the presence of chemical disorder in YbMgGaO$_4$  with randomly mixed occupancies of Mg$^{2+}$ and Ga$^{2+}$ at the same Wyckoff position \cite{ZhuPRL2017, ParkerPRB2018, ZhangPRX2018}. The closely related triangular lattice compound, YbZnGaO$_4$ attracted attention as well, but the chemical disorder clouded the interpretations \cite{MaPRL2018}. Intrigued by these results, a new ytterbium-based triangular family, known as chalcogenides (Na,Li)YbX$_2$ (X = O, S and Se), was  predicted to realize a disorder free QSL ground state \cite{BaenitzPRB2018, ranjith2019field,  BordelonNature2019, DingPRB2019, LiuChinLett2018, RanjithPRB2019, GuoPRM2020, BrodelonPRB2020}. The absence of significant site-disorder or magnetic ordering down to 70 mK was reported for these systems. However, most of the experiments were reported on polycrystalline samples as large sized single crystals of chalcogenides are challenging to grow. 

Recently, the disorder free ytterbium-based triangular lattice \bybo{} has also been identified. In \bybo{}, the magnetic ions Yb$^{3+}$-Yb$^{3+}$ are connected in a triangular lattice configuration in the $ab$ - plane (see supplementary materials). The intralayer distance between the Yb ions is $\simeq$ 5.4219 \AA, smaller than the interlayer distance of $\simeq$ 8.7311 \AA, therefore the magnetic coupling between nearest neighbor Yb$^{3+}$-Yb$^{3+}$ in the $ab$-plane is stronger than the coupling along the $c$-axis, making this system a two-dimensional triangular lattice. The nuclear magnetic resonance (NMR) experiment performed on this compound revealed absence of magnetic ordering or freezing down to 260 mK \cite{ZengPRB2019}. Further thermodynamics and $\mu$SR studies performed on polycrystalline \bybo{} sample show no signature of long range ordering within the experimental resolution limits \cite{Arxiv2021}. 

We successfully grew a centimeter sized single crystal sample of \bybo{} (see supplementary material). Our neutron diffraction measurements performed on single crystal sample, reveal the high quality of the grown crystal in bulk, while, pair distribution function (PDF) analyses performed on powder neutron diffraction data, confirm the absence of site mixing, chemical disorder, and any structural transition at low temperature (see supplementary materials). Having access to large high-quality single crystal sample, enables us to advance these studies in future and further investigate the properties of this system with techniques such as inelastic neutron scattering. Here, we report comprehensive thermodynamics studies of \bybo{} in the low temperature limit, revealing the absence of long range magnetic ordering down to 56 mK. We corroborate our experimental findings with theoretical analyses, and propose this compound may realize a fully quantum dipole-dipole interaction.

\begin{figure}[htbp]\centering\includegraphics[width= 8.8 cm]{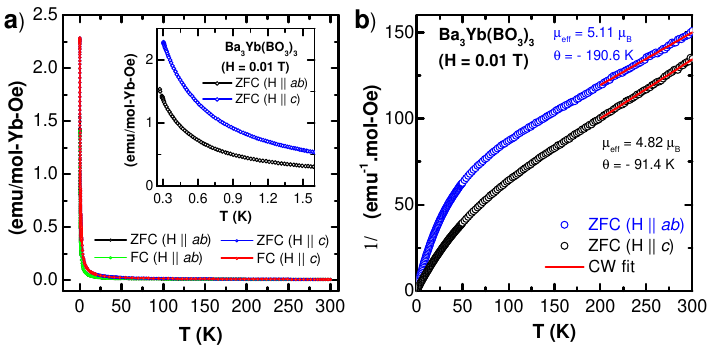}
	\caption{(a) Temperature dependent magnetic susceptibilities ($\chi_{ab}$ and $\chi_c$) obtained for \bybo{} crystal are shown. Directional magnetic anisotropy is more pronounced below 2 K, as shown in the inset. (b) Inverse magnetic susceptibility results and the Curie-Weiss fittings (solid red lines) obtained for high temperature regime are shown.}
	\label{MT}
\end{figure}

Temperature dependent magnetic susceptibility measurements were carried out on \bybo{} single crystal sample, in presence of an external magnetic field of H = 100 Oe along $ab$ - plane ($\chi_{ab}$, H $\parallel$ $ab$) and along $c$ - axis ($\chi_{c}$, H $\parallel$ $c$).  The results reveal no signs of magnetic ordering down to 290 mK (see Fig. \ref{MT}(a)). The susceptibility data shows directional magnetic anisotropy, where  $\chi_{c}$ is always higher in magnitude than $\chi_{ab}$, probably due to presence of anisotropic values of the Land\'e factor \cite{LiPRL2015,LiPRL2017,MaPRL2018}. In order to further confirm the observed directional magnetic anisotropy ($\chi_{c}$ $>$  $\chi_{ab}$), we performed similar magnetic susceptibility measurements on different single crystal samples of \bybo{} and concluded that the results are reproducible (see supplementary materials). The magnetic moment ($\chi_{c}$ $\sim$ 0.236 emu/mol-Yb-Oe and $\chi_{ab}$ $\sim$ 0.493 emu/mol-Yb-Oe at 2 K) obtained for our grown single crystal, agrees with the reported value  ($\sim$ 0.31 emu/mol-Yb-Oe at 2 K), which had been collected on polycrystalline sample \cite{GaoJAC2018}. 

In Fig. \ref{MT}(b), the inverse magnetic susceptibility data is fitted using the Curie-Weiss law: $1/\chi$ = $(T - \theta)/C$ ; where, `C' represents Curie constant, and `$\theta$' is Curie-Weiss temperature. The inverse susceptibility data (1/$\chi_{ab}$ and 1/$\chi_c$) is almost linear above 100 K, whereas a non-linear behavior is observed at low temperature. The Curie-Weiss fitting is done in the range from 300 K to 200 K. The effective moments ($\mu_{eff}$) are calculated from the fitted Curie constant `C'  at this temperature range. Considering that L = 3 and S = 1/2, (J = 7/2 and g = 8/7) for the Yb$^{3+}$ free ions, the expected magnetic moment ($\mu_{eff}$) at ambient temperature is 4.53 $\mu_{B}$ \cite{Blundell2001}. We note that the observed effective magnetic moments, obtained at high temperature range for both crystallographic directions of \bybo{} , are in agreement with the expected moment for Yb$^{3+}$ (see Fig. \ref{MT}(b)).   

The calculated effective magnetic moment ($\sim$ 5.11 $\mu_{B}$) along the $ab$ plane is found to be slightly higher than the moment observed ($\sim$ 4.82 $\mu_{B}$ ) along the crystallographic $c$-direction. The observed magnetic moments agree with the previously reported values obtained for Yb-based materials at ambient temperature \cite{BlotePhysica1969, CaoJPCM2009, SandersJPCM2017}. The Land\'e g-factor is also calculated from the Curie constant (C = $\frac{ng^2\mu_{B}^2J(J+1)} {3K_B}$ -- in which, `n' is the number of free spins per f.u., `$\mu_B$' is the Bohr magneton, and `K$_B$' is the Boltzmann constant), with the assumption that the Yb$^{3+}$ ions are free at ambient temperature. The obtained `g' values (g$_{ab}$ $\simeq$ 1.28 and g$_c$ $\simeq$ 1.21), with a slight anisotropy, agree with the expected value (g = 1.14).

A qualitative agreement of observed effective magnetic moments at ambient temperature and the expected values, in addition to the significant reduction of magnetic moment at low temperature, suggest that the Yb ions enter the Kramers doublet ground state at low temperature \cite{BlotePhysica1969, CaoJPCM2009, CavaPRM2019}. Considering  J$_{eff}$ = $\frac{1}{2}$ in \bybo{} \cite{ZengPRB2019}, the obtained g$_{ab}$ $\simeq$ 2.19 and g$_c$ $\simeq$ 3.38 values suggest a significant anisotropy at low temperature range. A similar anisotropy in `g' values at low temperature was also previously observed in other Yb-based triangular systems: YbMgGaO$_4$ (g$_{\perp}$ $\simeq$ 3.00 ; g$_{\parallel}$ $\simeq$ 3.82) and YbZnGaO$_4$ (g$_{\perp}$ $\simeq$ 3.17 ; g$_{\parallel}$ $\simeq$ 3.82) \cite{MaPRL2018}. The anisotropy detected in the susceptibility data is attributed to the presence of crystal-field effect, which further results in the emergence of a pronounced broad peak along H $\parallel$ \textit{ab} at low temperature range below 100 K (see Fig. \ref{MT}(b)). A similar directional anisotropy and broad peak are also reported for another Yb-based triangular system NaBaYb(BO$_3$)$_2$ \cite{CavaPRM2019}.

\begin{figure}[htbp]\centering\includegraphics[width= 8.5 cm]{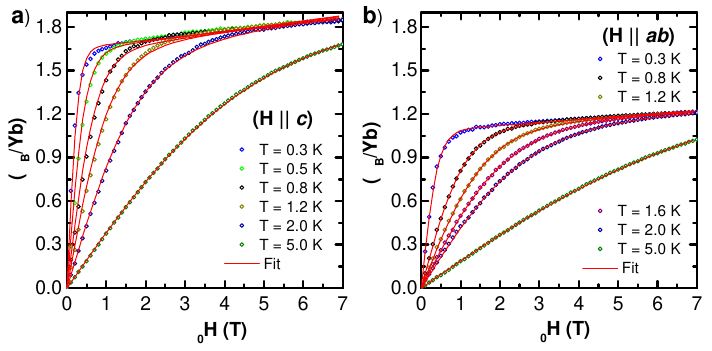}
	\caption{Isothermal magnetization data collected at different temperatures are plotted with lines fitted to a pure Van Vleck contribution and the Brillouin function of single-ion Zeeman energy, along crystallographic $c$-direction and (b) $ab$-plane of \bybo~.}
	\label{MH1}
\end{figure}

To understand the magnetic anisotropy and low temperature magnetism further, the field dependent isothermal magnetization measurements are carried out on \bybo{} single crystal samples. The M$_c$ and M$_{ab}$ results are collected at different temperatures (T = 0.3 K, 0.5 K, 0.8 K, 1.2 K, 1.6 K, 2.0 K, and 5.0 K) and up to 7 T in magnetic fields -- applied along \textit{c}-axis and \textit{ab}-plane of \bybo{} crystal, respectively (see Fig. \ref{MH1}). The data shows a nonlinear variation as a function of magnetic field. A significant rise in magnetization curves is observed for data collected at low temperature range (below 1 K). This enhancement in magnetization response is observed for both directions of \bybo{} crystal at fields below 1 T. The magnetization data seems to be saturated (or linearly increasing) at fields above $\sim$ 1 T along both directions. 

We fit the magnetization data to a pure Van Vleck contribution and the  Brillouin function of the single-ion Zeeman energy using the least squares method \cite{LiPRL2015}. The fitting agrees well with the data, as shown in Fig. \ref{MH1}. Adding exchange interactions does not improve the fitting in any significant way, indicating that the dominant energy scale in the data is the Zeeman energy, and that the exchange strengths are too small to be discerned from the data set. This is plausible given that the comparison material YbMgGaO$_4$, in which the Yb-Yb distance is 3.4 \AA, has estimated exchange parameters J$_z$ $\sim$ 0.98 K and J$\pm$ $\sim$ 0.9 K \cite{LiPRL2015}. Considering the large lattice spacing in \bybo{}, it is plausible that exchange coupling in this compound could be an order of magnitude smaller.

This observation suggests that the dominant interaction in \bybo{} might be the long-range dipole-dipole force. To estimate the dipole-dipole interaction energy, we take $\mu_{eff}=4.5\,\mu_B$ (magnetic moment of a free Yb$^{3+}$) and obtain the nearest-neighbor dipole energy of:

\begin{equation}
\label{dipole}
E_{\rm dipole} \approx - \frac{\mu_0}{4\pi r_0^3}. 2\mu_{eff}^2  \sim 0.0137\,\text{meV} \sim 0.16\,\text{K}.
\end{equation}

Such an exchange strength would negligibly affect the magnetization in the measured range, consistent with the fits above.  Indeed, in order to test the dominance of dipolar exchange of this magnitude, we must seek a measurement which is not overwhelmed by either thermal or Zeeman energies.  Precisely this is provided by a measurement of zero-field heat capacity, which we report below.

We note that the dipole-dipole interaction picture would lead to a significantly larger exchange interaction for the \textit{c}-axis oriented moments than for moments within the \textit{ab}-planes, $J_z \gg J_{xy}$: the out-of-plane components of nearest neighbor spins are highly frustrated due to the loop-like and long-range nature of the dipole-dipole interaction, giving a dominant antiferromagnetic interaction $J_z$. The in-plane exchange $J_{xy}$ between a given pair of spins, however, can be either antiferromagnetic or ferromagnetic depending on the details of the dipole orientations, and consequently the in-plane dipole interaction tends to be washed out at larger scales. 

\begin{figure*}[htbp]\centering\includegraphics[width = 16 cm]{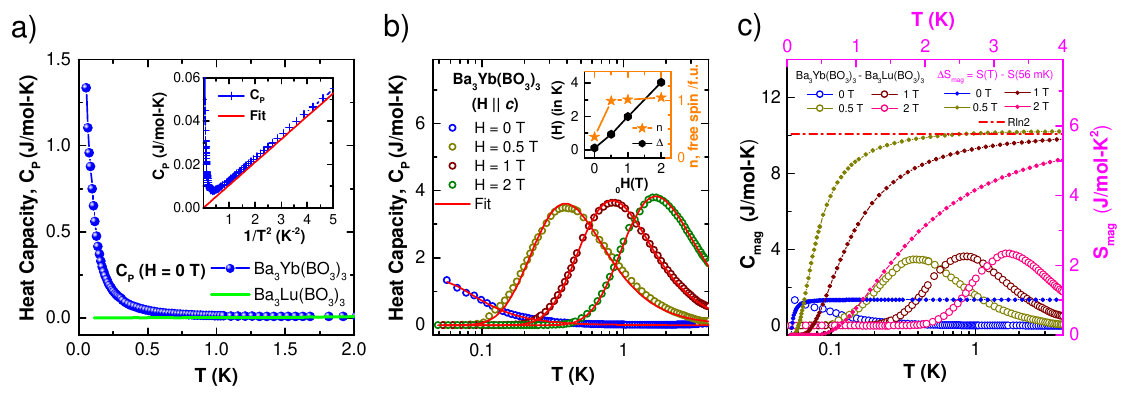}
	\caption{(a) The heat capacity results obtained for single crystalline \bybo{} and powder \blbo ~ samples are shown. Inset shows the heat capacity data (blue) and fit based on Eq.~\eqref{eq:cp_dipole} (red) as a function of $1/T^2$ at intermediate temperature range $0.5\,\text{K} \lesssim T \lesssim 4\,\text{K}$. (b) Heat capacity data plotted for various magnetic fields with fits (red solid lines) calculated for magnetic Schottky contribution, C$_{Sch}$(T, H)-- see Eq. \ref{Sch}. Inset shows the energy gap, $\Delta$(H) (left vertical- axis)) and `n' representing the number of free spins per f.u. (right vertical-axis) as function of applied magnetic fields. (c) Magnetic heat capacity $C_{mag}$ (left Y-axis, open symbols) of \bybo{} carried out under different magnetic fields are shown versus logarithmic temperature axis. The change in magnetic entropy ($\Delta$S$_{mag}$ = S(K)$-$S(56 mK), right vertical-axis, solid symbols) are plotted with linear temperature shown on the top-axis.}
	\label{HC1}
\end{figure*} 

The heat capacity measurements are carried out on a transparent and flat crystal specimen oriented perpendicular to the $c$-axis (see Fig. \ref{HC1}). The data is collected down to 56 mK at zero applied field. The results show almost zero heat capacity values near 2 K, which starts to increase below 1 K on cooling. For comparison, the heat capacity data is also collected on non-magnetic isostructure \blbo{}  powder sample, demonstrating almost zero values at temperatures below 2 K down to the base temperature of 100 mK -- suggesting almost zero phononic contribution in this temperature range. Furthermore, Yb$_2$O$_3$ is known to have Neel ordering around 2.3 K  \cite{MoonPR1968}. The absence of any anomaly in the heat capacity data of \bybo{} around 2.3 K suggests the absence of the magnetic Yb$_2$O$_3$ impurity in this sample and the high quality of the single crystal in general. Additionally, absence of any  anomaly in the measured heat capacity data down to 56 mK confirms that the system remains disordered down to the lowest accessible temperature. This is in agreement with our magnetic susceptibility data, showing no long-range magnetic ordering in this system. 

The enhanced heat capacity at low temperatures below $1$\,K would be expected due to weak but nonzero interactions between the moments in zero field, and this indeed gives a way to test the validity of the dipolar model. Within the dipolar model, the specific heat obtained from the high temperature expansion is:
\begin{equation}\label{eq:cp_dipole}
\begin{aligned}
    C_p &=  \frac{1}{2} \frac{\mu_B^4 S^4}{k_B T^2} \left(\frac{\mu_0}{4\pi}\right)^2  \sum_{\bm{r}\neq 0} \sum_{\mu\nu}g_{\mu\mu}^2 \left( \frac{r^\mu r^\nu - r^2 \delta^{\mu\nu}}{r^5}\right)^2 g_{\nu\nu}^2\\
    &=\frac{k_B}{|T|^2}0.000869 \,\,\,(\text{taking }g_c=3.38\text{ and }g_{ab}=2.2),
\end{aligned}
\end{equation}
where the first line gives the general expansion as a function of the $g$-factors, in which the sum over $\bm{r}$ runs over all the inter-spin vectors and $\mu,\nu$ sum over $x,y,z$. The second line makes use of the value of the $g$-factor obtained from the fitting in Fig.\ref{MH1}.

We compare this theoretical expression to experiment by plotting the zero field heat capacity measurement versus $1/T^2$ (see inset of Fig. \ref{HC1}(a)).  We observe a clear straight line, consistent with the {\em form} of Eq.~\eqref{eq:cp_dipole}.  A linear fit gives the coefficient:
\begin{equation}\label{eq:cp_fit}
    C_p = \frac{k_B}{|T|^2}0.00127.
\end{equation}
Comparing Eq.~(\ref{eq:cp_fit}) to Eq.~(\ref{eq:cp_dipole}), we see that the theoretical slope equals approximately 2/3 of the experimental one. This indicates that the majority of the exchange interaction is indeed dipolar, and is a strong validation of the dipolar picture.  It remains an open question to see whether the small discrepancy between the dipolar theory and experiment indicates a small super-exchange contribution, or may be related to the uncertainties of the heat capacity measurement. We now carry out further checks on the weakly dipolar nature of the system by measuring the heat capacity under several applied magnetic fields down to 56 mK.

In Fig. \ref{HC1}(b), the heat capacity data under several fields (0 T, 0.5 T, 1 T and 2 T) are shown below 4 K. The total heat capacity can be expressed as C$_P$(T, H) = C$_{lat}$ + C$_{Sch}$ (T, H). The first term is the lattice contribution (C$_{lat}$) which is almost zero in this temperature range, as explained above. The second term is the contribution from the two-level  Schottky anomaly given by:
\begin{equation}
\label{Sch}
C_{Sch}(T, H) = nR(\frac{\Delta}{T})^2\frac{exp(\Delta/T)}{(1 + exp(\Delta/T))^2}
\end{equation}
where, n is number of free spins per f.u. and $\Delta$ is the energy separation between two levels and `R' is ideal gas constant \cite{GangPRB2007,WangPRB2007,MuPRB2007,AdhikariPRB2019}. At low temperature, CEF excitation plays a crucial role for Yb$^{3+}$ based triangular systems as reported earlier \cite{LiPRB2016,DingPRB2019,RanjithPRB2019, LiPRL2017}. The combination of spin-orbit coupling and CEF excitation leads the eight multiplets ($^2F_{7/2}$) of Yb$^{3+}$ into four Kramers doublets \cite{DingPRB2019, LiPRL2017, RauPRL2016} and the low temperature properties can be described by spin $\frac{1}{2}$ Hamiltonian \cite{DingPRB2019}. This leads to Kramer doublet ground state of Yb$^{3+}$ (m$_J$ = $\pm$ $\frac{1}{2}$). In presence of magnetic field, the energy gap in Schottky anomaly can be expressed as $\Delta$(H) = g$\mu_B$H$_{eff}$/K$_B$; where, H$_{eff}$ = $\sqrt{H_0^2 + H^2}$ (H$_{eff}$ = H$_0$ is crystal field at zero external magnetic field) \cite{XiePhysica2012,WangPRB2007, MuPRB2007, AkbariJPCM2008}. The heat capacity data under various fields (H = 0 T, 0.5 T, 1 T and 2 T) are fitted using Schottky (C$_{Sch}$) contribution shown as red solid lines in Fig. \ref{HC1}(b). Under an applied field of 0.5 T, the heat capacity data shows a broad peak $\sim$ 0.4 K and, as expected, the peak temperature shifts towards higher temperatures with increasing magnetic field strength. 

The crystal field energy gap at zero field, $\Delta$(0), is estimated to be  0.113(2) K. The crystal field energy gap, $\Delta$(H), increases linearly with applied magnetic field (see inset of Fig. \ref{HC1}(b)). The origin of zero field energy gap could be related to the crystal field caused by small Yb$^{3+}$-Yb$^{3+}$ exchange interactions in the system. It is found that under 0.5 T of applied magnetic field, the energy levels split and most of the free spins ($\sim$ 99 $\%$) are excited to the higher energy levels. This agrees with our magnetization results (Fig. \ref{MH1}) -- showing a sharp enhancement below 1 T, with all spins polarized above 1 T. Similarly, the peak maxima for the heat capacity data remain unchanged, suggesting that $\sim$ 100 \% of the spins are free for 1 T and 2 T fields (see inset of Fig. \ref{HC1}(b)). Shifting of the peak position with higher fields suggests that there is a Zeeman splitting of the energy gap. The Land\'e `g'-factor is also calculated from the energy gap according to: $\Delta$(H) = g$\mu_B$H$_{eff}$/K$_B$. $\Delta$(H) is found to have a value of $\simeq$ 3.02(5) for fields below 2 T. As discussed above, this number is in agreement with the value (g$_c$ $\simeq$ 3.38) obtained from the magnetic measurements. 

To investigate the magnetic entropy at low temperatures, the non-magnetic \blbo{} heat capacity data is subtracted from \bybo{} data in order to remove the phononic contribution. The subtracted magnetic heat capacity C$_{mag}$ is plotted from 4 K to 56 mK as shown in Fig. \ref{HC1}(c). The C$_{mag}$ values obtained for \bybo{} for various fields are plotted (open symbols) on the left vertical-axis of Fig. \ref{HC1}(c) versus the logarithmic-temperature on the horizontal-axis. The magnetic entropy is calculated as $\Delta$S$_{mag}$  = $\int_{T_1}^{T_2}$ $\frac{C_{mag}}{T}dT$, by integrating  $\frac{C_{mag}}{T}$ from T$_1$ = 56 mK to T$_2$ = 4 K. The magnetic entropy obtained for different fields are plotted on the right vertical-axis with linear-temperature shown on top-axis (see Fig. \ref{HC1}(c)). Considering the effective spin (J$_{eff}$ = $\frac{1}{2}$) of Yb$^{3+}$ ions, the expected maximum entropy should be Rln(2J+1) = Rln2, entropy of a two-level system. Under zero field, the saturation value of magnetic entropy is found almost (1/5) of the value of Rln2 suggesting $\sim$ 80 \% of entropy remains below 56 mK. However, under applied magnetic field of 0.5 T, the calculated entropy saturates to Rln2, confirming the opening up of energy gap.  

In conclusion, large and high quality single crystals of triangular lattice \bybo{} were successfully grown using optical floating zone technique. A directional magnetic anisotropy is observed with anisotropic Land\'e g-factors. The magnetic and heat capacity data reveals the absence of any long-range magnetic ordering down to 56 mK. Field dependent heat capacity study suggests a magnetic ground state with a gap proportional to the applied field intensity. The magnetic entropy is related to the degrees of freedom of a two-level system with saturation value of Rln2 and effective spin $\frac{1}{2}$ for Yb ions. Based on the reported thermodynamics results, we propose that \bybo{} may realize an interesting example of the S = $\frac{1}{2}$ quantum dipole, in which the dominant interaction is the long range dipole-dipole interaction on the geometrically frustrated triangular lattice, with super-exchange interactions sub-dominant or negligible. Such pure dipolar systems are of fundamental interest. For example, it is known that for classical dipoles on the triangular lattice, the ground state is ferromagnetic for an infinite system, but may adopt complex spin textures in finite systems, depending on detailed geometry such as the aspect ratio \cite{PolitiPRB2006}. Overall, our study places \bybo{} in a similar category as the spin ice pyrochlores, albeit with some key differences: the spins in \bybo{} are XY-like rather than Ising, and hence fully quantum; the structure is triangular rather than pyrochlore. Future studies at the lowest temperatures would be interesting to explore quantum dipolar phenomena.

\section{Acknowledgements}
We are thankful to William Steinhardt for helpful discussions. The work at Duke University has been supported by William M. Fairbank Chair in Physics and NSF under grant DMR-1828348.  LB and CL were supported by the NSF CMMT program under Grant No. DMR-1818533. This research used resources at the High Flux Isotope Reactor and Spallation Neutron Source, a DOE Office of Science User Facility operated by the Oak Ridge National Laboratory.

\bibliographystyle{apsrev4-2}
\bibliography{BYBO}

\begin{thebibliography}{33}%
\makeatletter
\providecommand \@ifxundefined [1]{%
 \@ifx{#1\undefined}
}%
\providecommand \@ifnum [1]{%
 \ifnum #1\expandafter \@firstoftwo
 \else \expandafter \@secondoftwo
 \fi
}%
\providecommand \@ifx [1]{%
 \ifx #1\expandafter \@firstoftwo
 \else \expandafter \@secondoftwo
 \fi
}%
\providecommand \natexlab [1]{#1}%
\providecommand \enquote  [1]{``#1''}%
\providecommand \bibnamefont  [1]{#1}%
\providecommand \bibfnamefont [1]{#1}%
\providecommand \citenamefont [1]{#1}%
\providecommand \href@noop [0]{\@secondoftwo}%
\providecommand \href [0]{\begingroup \@sanitize@url \@href}%
\providecommand \@href[1]{\@@startlink{#1}\@@href}%
\providecommand \@@href[1]{\endgroup#1\@@endlink}%
\providecommand \@sanitize@url [0]{\catcode `\\12\catcode `\$12\catcode
  `\&12\catcode `\#12\catcode `\^12\catcode `\_12\catcode `\%12\relax}%
\providecommand \@@startlink[1]{}%
\providecommand \@@endlink[0]{}%
\providecommand \url  [0]{\begingroup\@sanitize@url \@url }%
\providecommand \@url [1]{\endgroup\@href {#1}{\urlprefix }}%
\providecommand \urlprefix  [0]{URL }%
\providecommand \Eprint [0]{\href }%
\providecommand \doibase [0]{https://doi.org/}%
\providecommand \selectlanguage [0]{\@gobble}%
\providecommand \bibinfo  [0]{\@secondoftwo}%
\providecommand \bibfield  [0]{\@secondoftwo}%
\providecommand \translation [1]{[#1]}%
\providecommand \BibitemOpen [0]{}%
\providecommand \bibitemStop [0]{}%
\providecommand \bibitemNoStop [0]{.\EOS\space}%
\providecommand \EOS [0]{\spacefactor3000\relax}%
\providecommand \BibitemShut  [1]{\csname bibitem#1\endcsname}%
\let\auto@bib@innerbib\@empty
\bibitem [{\citenamefont {Li}\ \emph {et~al.}(2015{\natexlab{a}})\citenamefont
  {Li}, \citenamefont {Chen}, \citenamefont {Tong}, \citenamefont {Pi},
  \citenamefont {Liu}, \citenamefont {Yang}, \citenamefont {Wang},\ and\
  \citenamefont {Zhang}}]{LiPRL2015}%
  \BibitemOpen
  \bibfield  {author} {\bibinfo {author} {\bibfnamefont {Y.}~\bibnamefont
  {Li}}, \bibinfo {author} {\bibfnamefont {G.}~\bibnamefont {Chen}}, \bibinfo
  {author} {\bibfnamefont {W.}~\bibnamefont {Tong}}, \bibinfo {author}
  {\bibfnamefont {L.}~\bibnamefont {Pi}}, \bibinfo {author} {\bibfnamefont
  {J.}~\bibnamefont {Liu}}, \bibinfo {author} {\bibfnamefont {Z.}~\bibnamefont
  {Yang}}, \bibinfo {author} {\bibfnamefont {X.}~\bibnamefont {Wang}},\ and\
  \bibinfo {author} {\bibfnamefont {Q.}~\bibnamefont {Zhang}},\ }\href
  {https://doi.org/10.1103/PhysRevLett.115.167203} {\bibfield  {journal}
  {\bibinfo  {journal} {Phys. Rev. Lett.}\ }\textbf {\bibinfo {volume} {115}},\
  \bibinfo {pages} {167203} (\bibinfo {year} {2015}{\natexlab{a}})}\BibitemShut
  {NoStop}%
\bibitem [{\citenamefont {Li}\ \emph {et~al.}(2015{\natexlab{b}})\citenamefont
  {Li}, \citenamefont {Liao}, \citenamefont {Zhang}, \citenamefont {Li},
  \citenamefont {Ling}, \citenamefont {Zhang}, \citenamefont {Zou},
  \citenamefont {Yang}, \citenamefont {Wang},\ and\ \citenamefont
  {Wu}}]{LiSciR2015}%
  \BibitemOpen
  \bibfield  {author} {\bibinfo {author} {\bibfnamefont {Y.}~\bibnamefont
  {Li}}, \bibinfo {author} {\bibfnamefont {H.}~\bibnamefont {Liao}}, \bibinfo
  {author} {\bibfnamefont {Z.}~\bibnamefont {Zhang}}, \bibinfo {author}
  {\bibfnamefont {S.}~\bibnamefont {Li}}, \bibinfo {author} {\bibfnamefont
  {L.}~\bibnamefont {Ling}}, \bibinfo {author} {\bibfnamefont {L.}~\bibnamefont
  {Zhang}}, \bibinfo {author} {\bibfnamefont {Y.}~\bibnamefont {Zou}}, \bibinfo
  {author} {\bibfnamefont {Z.}~\bibnamefont {Yang}}, \bibinfo {author}
  {\bibfnamefont {J.}~\bibnamefont {Wang}},\ and\ \bibinfo {author}
  {\bibfnamefont {Z.}~\bibnamefont {Wu}},\ }\href
  {https://doi.org/10.1038/srep16419} {\bibfield  {journal} {\bibinfo
  {journal} {Scientific Reports}\ }\textbf {\bibinfo {volume} {5}},\ \bibinfo
  {pages} {16419} (\bibinfo {year} {2015}{\natexlab{b}})}\BibitemShut {NoStop}%
\bibitem [{\citenamefont {Li}\ \emph {et~al.}(2016)\citenamefont {Li},
  \citenamefont {Wang},\ and\ \citenamefont {Chen}}]{LiPRB2016}%
  \BibitemOpen
  \bibfield  {author} {\bibinfo {author} {\bibfnamefont {Y.-D.}\ \bibnamefont
  {Li}}, \bibinfo {author} {\bibfnamefont {X.}~\bibnamefont {Wang}},\ and\
  \bibinfo {author} {\bibfnamefont {G.}~\bibnamefont {Chen}},\ }\href
  {https://doi.org/10.1103/PhysRevB.94.035107} {\bibfield  {journal} {\bibinfo
  {journal} {Phys. Rev. B}\ }\textbf {\bibinfo {volume} {94}},\ \bibinfo
  {pages} {035107} (\bibinfo {year} {2016})}\BibitemShut {NoStop}%
\bibitem [{\citenamefont {Li}\ \emph {et~al.}(2017)\citenamefont {Li},
  \citenamefont {Adroja}, \citenamefont {Bewley}, \citenamefont {Voneshen},
  \citenamefont {Tsirlin}, \citenamefont {Gegenwart},\ and\ \citenamefont
  {Zhang}}]{LiPRL2017}%
  \BibitemOpen
  \bibfield  {author} {\bibinfo {author} {\bibfnamefont {Y.}~\bibnamefont
  {Li}}, \bibinfo {author} {\bibfnamefont {D.}~\bibnamefont {Adroja}}, \bibinfo
  {author} {\bibfnamefont {R.~I.}\ \bibnamefont {Bewley}}, \bibinfo {author}
  {\bibfnamefont {D.}~\bibnamefont {Voneshen}}, \bibinfo {author}
  {\bibfnamefont {A.~A.}\ \bibnamefont {Tsirlin}}, \bibinfo {author}
  {\bibfnamefont {P.}~\bibnamefont {Gegenwart}},\ and\ \bibinfo {author}
  {\bibfnamefont {Q.}~\bibnamefont {Zhang}},\ }\href
  {https://doi.org/10.1103/PhysRevLett.118.107202} {\bibfield  {journal}
  {\bibinfo  {journal} {Phys. Rev. Lett.}\ }\textbf {\bibinfo {volume} {118}},\
  \bibinfo {pages} {107202} (\bibinfo {year} {2017})}\BibitemShut {NoStop}%
\bibitem [{\citenamefont {Zhu}\ \emph {et~al.}(2017)\citenamefont {Zhu},
  \citenamefont {Maksimov}, \citenamefont {White},\ and\ \citenamefont
  {Chernyshev}}]{ZhuPRL2017}%
  \BibitemOpen
  \bibfield  {author} {\bibinfo {author} {\bibfnamefont {Z.}~\bibnamefont
  {Zhu}}, \bibinfo {author} {\bibfnamefont {P.~A.}\ \bibnamefont {Maksimov}},
  \bibinfo {author} {\bibfnamefont {S.~R.}\ \bibnamefont {White}},\ and\
  \bibinfo {author} {\bibfnamefont {A.~L.}\ \bibnamefont {Chernyshev}},\ }\href
  {https://doi.org/10.1103/PhysRevLett.119.157201} {\bibfield  {journal}
  {\bibinfo  {journal} {Phys. Rev. Lett.}\ }\textbf {\bibinfo {volume} {119}},\
  \bibinfo {pages} {157201} (\bibinfo {year} {2017})}\BibitemShut {NoStop}%
\bibitem [{\citenamefont {Parker}\ and\ \citenamefont
  {Balents}(2018)}]{ParkerPRB2018}%
  \BibitemOpen
  \bibfield  {author} {\bibinfo {author} {\bibfnamefont {E.}~\bibnamefont
  {Parker}}\ and\ \bibinfo {author} {\bibfnamefont {L.}~\bibnamefont
  {Balents}},\ }\href {https://doi.org/10.1103/PhysRevB.97.184413} {\bibfield
  {journal} {\bibinfo  {journal} {Phys. Rev. B}\ }\textbf {\bibinfo {volume}
  {97}},\ \bibinfo {pages} {184413} (\bibinfo {year} {2018})}\BibitemShut
  {NoStop}%
\bibitem [{\citenamefont {Zhang}\ \emph {et~al.}(2018)\citenamefont {Zhang},
  \citenamefont {Mahmood}, \citenamefont {Daum}, \citenamefont {Dun},
  \citenamefont {Paddison}, \citenamefont {Laurita}, \citenamefont {Hong},
  \citenamefont {Zhou}, \citenamefont {Armitage},\ and\ \citenamefont
  {Mourigal}}]{ZhangPRX2018}%
  \BibitemOpen
  \bibfield  {author} {\bibinfo {author} {\bibfnamefont {X.}~\bibnamefont
  {Zhang}}, \bibinfo {author} {\bibfnamefont {F.}~\bibnamefont {Mahmood}},
  \bibinfo {author} {\bibfnamefont {M.}~\bibnamefont {Daum}}, \bibinfo {author}
  {\bibfnamefont {Z.}~\bibnamefont {Dun}}, \bibinfo {author} {\bibfnamefont
  {J.~A.~M.}\ \bibnamefont {Paddison}}, \bibinfo {author} {\bibfnamefont
  {N.~J.}\ \bibnamefont {Laurita}}, \bibinfo {author} {\bibfnamefont
  {T.}~\bibnamefont {Hong}}, \bibinfo {author} {\bibfnamefont {H.}~\bibnamefont
  {Zhou}}, \bibinfo {author} {\bibfnamefont {N.~P.}\ \bibnamefont {Armitage}},\
  and\ \bibinfo {author} {\bibfnamefont {M.}~\bibnamefont {Mourigal}},\ }\href
  {https://doi.org/10.1103/PhysRevX.8.031001} {\bibfield  {journal} {\bibinfo
  {journal} {Phys. Rev. X}\ }\textbf {\bibinfo {volume} {8}},\ \bibinfo {pages}
  {031001} (\bibinfo {year} {2018})}\BibitemShut {NoStop}%
\bibitem [{\citenamefont {Ma}\ \emph {et~al.}(2018)\citenamefont {Ma},
  \citenamefont {Wang}, \citenamefont {Dong}, \citenamefont {Zhang},
  \citenamefont {Li}, \citenamefont {Zheng}, \citenamefont {Yu}, \citenamefont
  {Wang}, \citenamefont {Che}, \citenamefont {Ran}, \citenamefont {Bao},
  \citenamefont {Cai}, \citenamefont {\ifmmode~\check{C}\else
  \v{C}\fi{}erm\'ak}, \citenamefont {Schneidewind}, \citenamefont {Yano},
  \citenamefont {Gardner}, \citenamefont {Lu}, \citenamefont {Yu},
  \citenamefont {Liu}, \citenamefont {Li}, \citenamefont {Li},\ and\
  \citenamefont {Wen}}]{MaPRL2018}%
  \BibitemOpen
  \bibfield  {author} {\bibinfo {author} {\bibfnamefont {Z.}~\bibnamefont
  {Ma}}, \bibinfo {author} {\bibfnamefont {J.}~\bibnamefont {Wang}}, \bibinfo
  {author} {\bibfnamefont {Z.-Y.}\ \bibnamefont {Dong}}, \bibinfo {author}
  {\bibfnamefont {J.}~\bibnamefont {Zhang}}, \bibinfo {author} {\bibfnamefont
  {S.}~\bibnamefont {Li}}, \bibinfo {author} {\bibfnamefont {S.-H.}\
  \bibnamefont {Zheng}}, \bibinfo {author} {\bibfnamefont {Y.}~\bibnamefont
  {Yu}}, \bibinfo {author} {\bibfnamefont {W.}~\bibnamefont {Wang}}, \bibinfo
  {author} {\bibfnamefont {L.}~\bibnamefont {Che}}, \bibinfo {author}
  {\bibfnamefont {K.}~\bibnamefont {Ran}}, \bibinfo {author} {\bibfnamefont
  {S.}~\bibnamefont {Bao}}, \bibinfo {author} {\bibfnamefont {Z.}~\bibnamefont
  {Cai}}, \bibinfo {author} {\bibfnamefont {P.}~\bibnamefont
  {\ifmmode~\check{C}\else \v{C}\fi{}erm\'ak}}, \bibinfo {author}
  {\bibfnamefont {A.}~\bibnamefont {Schneidewind}}, \bibinfo {author}
  {\bibfnamefont {S.}~\bibnamefont {Yano}}, \bibinfo {author} {\bibfnamefont
  {J.~S.}\ \bibnamefont {Gardner}}, \bibinfo {author} {\bibfnamefont
  {X.}~\bibnamefont {Lu}}, \bibinfo {author} {\bibfnamefont {S.-L.}\
  \bibnamefont {Yu}}, \bibinfo {author} {\bibfnamefont {J.-M.}\ \bibnamefont
  {Liu}}, \bibinfo {author} {\bibfnamefont {S.}~\bibnamefont {Li}}, \bibinfo
  {author} {\bibfnamefont {J.-X.}\ \bibnamefont {Li}},\ and\ \bibinfo {author}
  {\bibfnamefont {J.}~\bibnamefont {Wen}},\ }\href
  {https://doi.org/10.1103/PhysRevLett.120.087201} {\bibfield  {journal}
  {\bibinfo  {journal} {Phys. Rev. Lett.}\ }\textbf {\bibinfo {volume} {120}},\
  \bibinfo {pages} {087201} (\bibinfo {year} {2018})}\BibitemShut {NoStop}%
\bibitem [{\citenamefont {Baenitz}\ \emph {et~al.}(2018)\citenamefont
  {Baenitz}, \citenamefont {Schlender}, \citenamefont {Sichelschmidt},
  \citenamefont {Onykiienko}, \citenamefont {Zangeneh}, \citenamefont
  {Ranjith}, \citenamefont {Sarkar}, \citenamefont {Hozoi}, \citenamefont
  {Walker}, \citenamefont {Orain}, \citenamefont {Yasuoka}, \citenamefont
  {van~den Brink}, \citenamefont {Klauss}, \citenamefont {Inosov},\ and\
  \citenamefont {Doert}}]{BaenitzPRB2018}%
  \BibitemOpen
  \bibfield  {author} {\bibinfo {author} {\bibfnamefont {M.}~\bibnamefont
  {Baenitz}}, \bibinfo {author} {\bibfnamefont {P.}~\bibnamefont {Schlender}},
  \bibinfo {author} {\bibfnamefont {J.}~\bibnamefont {Sichelschmidt}}, \bibinfo
  {author} {\bibfnamefont {Y.~A.}\ \bibnamefont {Onykiienko}}, \bibinfo
  {author} {\bibfnamefont {Z.}~\bibnamefont {Zangeneh}}, \bibinfo {author}
  {\bibfnamefont {K.~M.}\ \bibnamefont {Ranjith}}, \bibinfo {author}
  {\bibfnamefont {R.}~\bibnamefont {Sarkar}}, \bibinfo {author} {\bibfnamefont
  {L.}~\bibnamefont {Hozoi}}, \bibinfo {author} {\bibfnamefont {H.~C.}\
  \bibnamefont {Walker}}, \bibinfo {author} {\bibfnamefont {J.-C.}\
  \bibnamefont {Orain}}, \bibinfo {author} {\bibfnamefont {H.}~\bibnamefont
  {Yasuoka}}, \bibinfo {author} {\bibfnamefont {J.}~\bibnamefont {van~den
  Brink}}, \bibinfo {author} {\bibfnamefont {H.~H.}\ \bibnamefont {Klauss}},
  \bibinfo {author} {\bibfnamefont {D.~S.}\ \bibnamefont {Inosov}},\ and\
  \bibinfo {author} {\bibfnamefont {T.}~\bibnamefont {Doert}},\ }\href
  {https://doi.org/10.1103/PhysRevB.98.220409} {\bibfield  {journal} {\bibinfo
  {journal} {Phys. Rev. B}\ }\textbf {\bibinfo {volume} {98}},\ \bibinfo
  {pages} {220409} (\bibinfo {year} {2018})}\BibitemShut {NoStop}%
\bibitem [{\citenamefont {Bordelon}\ \emph {et~al.}(2019)\citenamefont
  {Bordelon}, \citenamefont {Kenney},\ and\ \citenamefont
  {Liu}}]{BordelonNature2019}%
  \BibitemOpen
  \bibfield  {author} {\bibinfo {author} {\bibfnamefont {M.}~\bibnamefont
  {Bordelon}}, \bibinfo {author} {\bibfnamefont {E.}~\bibnamefont {Kenney}},\
  and\ \bibinfo {author} {\bibfnamefont {C.}~\bibnamefont {Liu}},\ }\href
  {https://doi.org/10.1038/s41567-019-0594-5} {\bibfield  {journal} {\bibinfo
  {journal} {Nature Phys}\ }\textbf {\bibinfo {volume} {15}},\ \bibinfo {pages}
  {1058} (\bibinfo {year} {2019})}\BibitemShut {NoStop}%
\bibitem [{\citenamefont {Ding}\ \emph {et~al.}(2019)\citenamefont {Ding},
  \citenamefont {Manuel}, \citenamefont {Bachus}, \citenamefont {Gru\ss{}ler},
  \citenamefont {Gegenwart}, \citenamefont {Singleton}, \citenamefont
  {Johnson}, \citenamefont {Walker}, \citenamefont {Adroja}, \citenamefont
  {Hillier},\ and\ \citenamefont {Tsirlin}}]{DingPRB2019}%
  \BibitemOpen
  \bibfield  {author} {\bibinfo {author} {\bibfnamefont {L.}~\bibnamefont
  {Ding}}, \bibinfo {author} {\bibfnamefont {P.}~\bibnamefont {Manuel}},
  \bibinfo {author} {\bibfnamefont {S.}~\bibnamefont {Bachus}}, \bibinfo
  {author} {\bibfnamefont {F.}~\bibnamefont {Gru\ss{}ler}}, \bibinfo {author}
  {\bibfnamefont {P.}~\bibnamefont {Gegenwart}}, \bibinfo {author}
  {\bibfnamefont {J.}~\bibnamefont {Singleton}}, \bibinfo {author}
  {\bibfnamefont {R.~D.}\ \bibnamefont {Johnson}}, \bibinfo {author}
  {\bibfnamefont {H.~C.}\ \bibnamefont {Walker}}, \bibinfo {author}
  {\bibfnamefont {D.~T.}\ \bibnamefont {Adroja}}, \bibinfo {author}
  {\bibfnamefont {A.~D.}\ \bibnamefont {Hillier}},\ and\ \bibinfo {author}
  {\bibfnamefont {A.~A.}\ \bibnamefont {Tsirlin}},\ }\href
  {https://doi.org/10.1103/PhysRevB.100.144432} {\bibfield  {journal} {\bibinfo
   {journal} {Phys. Rev. B}\ }\textbf {\bibinfo {volume} {100}},\ \bibinfo
  {pages} {144432} (\bibinfo {year} {2019})}\BibitemShut {NoStop}%
\bibitem [{\citenamefont {Liu}\ \emph {et~al.}(2018)\citenamefont {Liu},
  \citenamefont {Zhang}, \citenamefont {Ji}, \citenamefont {Liu}, \citenamefont
  {Li}, \citenamefont {Wang}, \citenamefont {Lei}, \citenamefont {Chen},\ and\
  \citenamefont {Zhang}}]{LiuChinLett2018}%
  \BibitemOpen
  \bibfield  {author} {\bibinfo {author} {\bibfnamefont {W.}~\bibnamefont
  {Liu}}, \bibinfo {author} {\bibfnamefont {Z.}~\bibnamefont {Zhang}}, \bibinfo
  {author} {\bibfnamefont {J.}~\bibnamefont {Ji}}, \bibinfo {author}
  {\bibfnamefont {Y.}~\bibnamefont {Liu}}, \bibinfo {author} {\bibfnamefont
  {J.}~\bibnamefont {Li}}, \bibinfo {author} {\bibfnamefont {X.}~\bibnamefont
  {Wang}}, \bibinfo {author} {\bibfnamefont {H.}~\bibnamefont {Lei}}, \bibinfo
  {author} {\bibfnamefont {G.}~\bibnamefont {Chen}},\ and\ \bibinfo {author}
  {\bibfnamefont {Q.}~\bibnamefont {Zhang}},\ }\href
  {https://doi.org/10.1088/0256-307x/35/11/117501} {\bibfield  {journal}
  {\bibinfo  {journal} {Chinese Physics Letters}\ }\textbf {\bibinfo {volume}
  {35}},\ \bibinfo {pages} {117501} (\bibinfo {year} {2018})}\BibitemShut
  {NoStop}%
\bibitem [{\citenamefont {Ranjith}\ \emph
  {et~al.}(2019{\natexlab{a}})\citenamefont {Ranjith}, \citenamefont {Luther},
  \citenamefont {Reimann}, \citenamefont {Schmidt}, \citenamefont {Schlender},
  \citenamefont {Sichelschmidt}, \citenamefont {Yasuoka}, \citenamefont
  {Strydom}, \citenamefont {Skourski}, \citenamefont {Wosnitza}, \citenamefont
  {K\"uhne}, \citenamefont {Doert},\ and\ \citenamefont
  {Baenitz}}]{RanjithPRB2019}%
  \BibitemOpen
  \bibfield  {author} {\bibinfo {author} {\bibfnamefont {K.~M.}\ \bibnamefont
  {Ranjith}}, \bibinfo {author} {\bibfnamefont {S.}~\bibnamefont {Luther}},
  \bibinfo {author} {\bibfnamefont {T.}~\bibnamefont {Reimann}}, \bibinfo
  {author} {\bibfnamefont {B.}~\bibnamefont {Schmidt}}, \bibinfo {author}
  {\bibfnamefont {P.}~\bibnamefont {Schlender}}, \bibinfo {author}
  {\bibfnamefont {J.}~\bibnamefont {Sichelschmidt}}, \bibinfo {author}
  {\bibfnamefont {H.}~\bibnamefont {Yasuoka}}, \bibinfo {author} {\bibfnamefont
  {A.~M.}\ \bibnamefont {Strydom}}, \bibinfo {author} {\bibfnamefont
  {Y.}~\bibnamefont {Skourski}}, \bibinfo {author} {\bibfnamefont
  {J.}~\bibnamefont {Wosnitza}}, \bibinfo {author} {\bibfnamefont
  {H.}~\bibnamefont {K\"uhne}}, \bibinfo {author} {\bibfnamefont
  {T.}~\bibnamefont {Doert}},\ and\ \bibinfo {author} {\bibfnamefont
  {M.}~\bibnamefont {Baenitz}},\ }\href
  {https://doi.org/10.1103/PhysRevB.100.224417} {\bibfield  {journal} {\bibinfo
   {journal} {Phys. Rev. B}\ }\textbf {\bibinfo {volume} {100}},\ \bibinfo
  {pages} {224417} (\bibinfo {year} {2019}{\natexlab{a}})}\BibitemShut
  {NoStop}%
\bibitem [{\citenamefont {Guo}\ \emph {et~al.}(2020)\citenamefont {Guo},
  \citenamefont {Zhao}, \citenamefont {Ohira-Kawamura}, \citenamefont {Ling},
  \citenamefont {Wang}, \citenamefont {He}, \citenamefont {Nakajima},
  \citenamefont {Li},\ and\ \citenamefont {Zhang}}]{GuoPRM2020}%
  \BibitemOpen
  \bibfield  {author} {\bibinfo {author} {\bibfnamefont {J.}~\bibnamefont
  {Guo}}, \bibinfo {author} {\bibfnamefont {X.}~\bibnamefont {Zhao}}, \bibinfo
  {author} {\bibfnamefont {S.}~\bibnamefont {Ohira-Kawamura}}, \bibinfo
  {author} {\bibfnamefont {L.}~\bibnamefont {Ling}}, \bibinfo {author}
  {\bibfnamefont {J.}~\bibnamefont {Wang}}, \bibinfo {author} {\bibfnamefont
  {L.}~\bibnamefont {He}}, \bibinfo {author} {\bibfnamefont {K.}~\bibnamefont
  {Nakajima}}, \bibinfo {author} {\bibfnamefont {B.}~\bibnamefont {Li}},\ and\
  \bibinfo {author} {\bibfnamefont {Z.}~\bibnamefont {Zhang}},\ }\href
  {https://doi.org/10.1103/PhysRevMaterials.4.064410} {\bibfield  {journal}
  {\bibinfo  {journal} {Phys. Rev. Materials}\ }\textbf {\bibinfo {volume}
  {4}},\ \bibinfo {pages} {064410} (\bibinfo {year} {2020})}\BibitemShut
  {NoStop}%
\bibitem [{\citenamefont {Bordelon}\ \emph {et~al.}(2020)\citenamefont
  {Bordelon}, \citenamefont {Liu}, \citenamefont {Posthuma}, \citenamefont
  {Sarte}, \citenamefont {Butch}, \citenamefont {Pajerowski}, \citenamefont
  {Banerjee}, \citenamefont {Balents},\ and\ \citenamefont
  {Wilson}}]{BrodelonPRB2020}%
  \BibitemOpen
  \bibfield  {author} {\bibinfo {author} {\bibfnamefont {M.~M.}\ \bibnamefont
  {Bordelon}}, \bibinfo {author} {\bibfnamefont {C.}~\bibnamefont {Liu}},
  \bibinfo {author} {\bibfnamefont {L.}~\bibnamefont {Posthuma}}, \bibinfo
  {author} {\bibfnamefont {P.~M.}\ \bibnamefont {Sarte}}, \bibinfo {author}
  {\bibfnamefont {N.~P.}\ \bibnamefont {Butch}}, \bibinfo {author}
  {\bibfnamefont {D.~M.}\ \bibnamefont {Pajerowski}}, \bibinfo {author}
  {\bibfnamefont {A.}~\bibnamefont {Banerjee}}, \bibinfo {author}
  {\bibfnamefont {L.}~\bibnamefont {Balents}},\ and\ \bibinfo {author}
  {\bibfnamefont {S.~D.}\ \bibnamefont {Wilson}},\ }\href
  {https://doi.org/10.1103/PhysRevB.101.224427} {\bibfield  {journal} {\bibinfo
   {journal} {Phys. Rev. B}\ }\textbf {\bibinfo {volume} {101}},\ \bibinfo
  {pages} {224427} (\bibinfo {year} {2020})}\BibitemShut {NoStop}%
\bibitem [{\citenamefont {Ranjith}\ \emph
  {et~al.}(2019{\natexlab{b}})\citenamefont {Ranjith}, \citenamefont
  {Dmytriieva}, \citenamefont {Khim}, \citenamefont {Sichelschmidt},
  \citenamefont {Luther}, \citenamefont {Ehlers}, \citenamefont {Yasuoka},
  \citenamefont {Wosnitza}, \citenamefont {Tsirlin}, \citenamefont {K{\"u}hne}
  \emph {et~al.}}]{ranjith2019field}%
  \BibitemOpen
  \bibfield  {author} {\bibinfo {author} {\bibfnamefont {K.}~\bibnamefont
  {Ranjith}}, \bibinfo {author} {\bibfnamefont {D.}~\bibnamefont {Dmytriieva}},
  \bibinfo {author} {\bibfnamefont {S.}~\bibnamefont {Khim}}, \bibinfo {author}
  {\bibfnamefont {J.}~\bibnamefont {Sichelschmidt}}, \bibinfo {author}
  {\bibfnamefont {S.}~\bibnamefont {Luther}}, \bibinfo {author} {\bibfnamefont
  {D.}~\bibnamefont {Ehlers}}, \bibinfo {author} {\bibfnamefont
  {H.}~\bibnamefont {Yasuoka}}, \bibinfo {author} {\bibfnamefont
  {J.}~\bibnamefont {Wosnitza}}, \bibinfo {author} {\bibfnamefont {A.~A.}\
  \bibnamefont {Tsirlin}}, \bibinfo {author} {\bibfnamefont {H.}~\bibnamefont
  {K{\"u}hne}}, \emph {et~al.},\ }\href@noop {} {\bibfield  {journal} {\bibinfo
   {journal} {Physical Review B}\ }\textbf {\bibinfo {volume} {99}},\ \bibinfo
  {pages} {180401} (\bibinfo {year} {2019}{\natexlab{b}})}\BibitemShut
  {NoStop}%
\bibitem [{\citenamefont {Zeng}\ \emph {et~al.}(2020)\citenamefont {Zeng},
  \citenamefont {Ma}, \citenamefont {Gao}, \citenamefont {Tian}, \citenamefont
  {Ling},\ and\ \citenamefont {Pi}}]{ZengPRB2019}%
  \BibitemOpen
  \bibfield  {author} {\bibinfo {author} {\bibfnamefont {K.~Y.}\ \bibnamefont
  {Zeng}}, \bibinfo {author} {\bibfnamefont {L.}~\bibnamefont {Ma}}, \bibinfo
  {author} {\bibfnamefont {Y.~X.}\ \bibnamefont {Gao}}, \bibinfo {author}
  {\bibfnamefont {Z.~M.}\ \bibnamefont {Tian}}, \bibinfo {author}
  {\bibfnamefont {L.~S.}\ \bibnamefont {Ling}},\ and\ \bibinfo {author}
  {\bibfnamefont {L.}~\bibnamefont {Pi}},\ }\href
  {https://doi.org/10.1103/PhysRevB.102.045149} {\bibfield  {journal} {\bibinfo
   {journal} {Phys. Rev. B}\ }\textbf {\bibinfo {volume} {102}},\ \bibinfo
  {pages} {045149} (\bibinfo {year} {2020})}\BibitemShut {NoStop}%
\bibitem [{\citenamefont {Cho}\ \emph {et~al.}(2021)\citenamefont {Cho},
  \citenamefont {Blundell}, \citenamefont {Shiroka}, \citenamefont
  {MacFarquharson}, \citenamefont {Prabhakaran},\ and\ \citenamefont
  {Coldea}}]{Arxiv2021}%
  \BibitemOpen
  \bibfield  {author} {\bibinfo {author} {\bibfnamefont {H.}~\bibnamefont
  {Cho}}, \bibinfo {author} {\bibfnamefont {S.~J.}\ \bibnamefont {Blundell}},
  \bibinfo {author} {\bibfnamefont {T.}~\bibnamefont {Shiroka}}, \bibinfo
  {author} {\bibfnamefont {K.}~\bibnamefont {MacFarquharson}}, \bibinfo
  {author} {\bibfnamefont {D.}~\bibnamefont {Prabhakaran}},\ and\ \bibinfo
  {author} {\bibfnamefont {R.}~\bibnamefont {Coldea}},\ }\href@noop {}
  {\bibinfo {title} {Studies on novel yb-based candidate triangular quantum
  antiferromagnets: Ba3ybb3o9 and ba3ybb9o18}} (\bibinfo {year} {2021}),\
  \Eprint {https://arxiv.org/abs/2104.01005} {arXiv:2104.01005
  [cond-mat.str-el]} \BibitemShut {NoStop}%
\bibitem [{\citenamefont {Gao}\ \emph {et~al.}(2018)\citenamefont {Gao},
  \citenamefont {Xu}, \citenamefont {Tian},\ and\ \citenamefont
  {Yuan}}]{GaoJAC2018}%
  \BibitemOpen
  \bibfield  {author} {\bibinfo {author} {\bibfnamefont {Y.}~\bibnamefont
  {Gao}}, \bibinfo {author} {\bibfnamefont {L.}~\bibnamefont {Xu}}, \bibinfo
  {author} {\bibfnamefont {Z.}~\bibnamefont {Tian}},\ and\ \bibinfo {author}
  {\bibfnamefont {S.}~\bibnamefont {Yuan}},\ }\href
  {https://doi.org/https://doi.org/10.1016/j.jallcom.2018.02.110} {\bibfield
  {journal} {\bibinfo  {journal} {Journal of Alloys and Compounds}\ }\textbf
  {\bibinfo {volume} {745}},\ \bibinfo {pages} {396 } (\bibinfo {year}
  {2018})}\BibitemShut {NoStop}%
\bibitem [{\citenamefont {Blundell}(2001)}]{Blundell2001}%
  \BibitemOpen
  \bibfield  {author} {\bibinfo {author} {\bibfnamefont {S.}~\bibnamefont
  {Blundell}},\ }\href {https://books.google.co.in/books?id=KqdQrgEACAAJ}
  {\emph {\bibinfo {title} {{Magnetism in Condensed Matter}}}},\ Oxford Master
  Series in Condensed Matter Physics 4\ (\bibinfo  {publisher} {OUP Oxford},\
  \bibinfo {year} {2001})\BibitemShut {NoStop}%
\bibitem [{\citenamefont {Blöte}\ \emph {et~al.}(1969)\citenamefont {Blöte},
  \citenamefont {Wielinga},\ and\ \citenamefont {Huiskamp}}]{BlotePhysica1969}%
  \BibitemOpen
  \bibfield  {author} {\bibinfo {author} {\bibfnamefont {H.}~\bibnamefont
  {Blöte}}, \bibinfo {author} {\bibfnamefont {R.}~\bibnamefont {Wielinga}},\
  and\ \bibinfo {author} {\bibfnamefont {W.}~\bibnamefont {Huiskamp}},\ }\href
  {https://doi.org/https://doi.org/10.1016/0031-8914(69)90187-6} {\bibfield
  {journal} {\bibinfo  {journal} {Physica}\ }\textbf {\bibinfo {volume} {43}},\
  \bibinfo {pages} {549 } (\bibinfo {year} {1969})}\BibitemShut {NoStop}%
\bibitem [{\citenamefont {Cao}\ \emph {et~al.}(2009)\citenamefont {Cao},
  \citenamefont {Gukasov}, \citenamefont {Mirebeau},\ and\ \citenamefont
  {Bonville}}]{CaoJPCM2009}%
  \BibitemOpen
  \bibfield  {author} {\bibinfo {author} {\bibfnamefont {H.~B.}\ \bibnamefont
  {Cao}}, \bibinfo {author} {\bibfnamefont {A.}~\bibnamefont {Gukasov}},
  \bibinfo {author} {\bibfnamefont {I.}~\bibnamefont {Mirebeau}},\ and\
  \bibinfo {author} {\bibfnamefont {P.}~\bibnamefont {Bonville}},\ }\href
  {https://doi.org/10.1088/0953-8984/21/49/492202} {\bibfield  {journal}
  {\bibinfo  {journal} {Journal of Physics: Condensed Matter}\ }\textbf
  {\bibinfo {volume} {21}},\ \bibinfo {pages} {492202} (\bibinfo {year}
  {2009})}\BibitemShut {NoStop}%
\bibitem [{\citenamefont {Sanders}\ \emph {et~al.}(2017)\citenamefont
  {Sanders}, \citenamefont {Cevallos},\ and\ \citenamefont
  {Cava}}]{SandersJPCM2017}%
  \BibitemOpen
  \bibfield  {author} {\bibinfo {author} {\bibfnamefont {M.~B.}\ \bibnamefont
  {Sanders}}, \bibinfo {author} {\bibfnamefont {F.~A.}\ \bibnamefont
  {Cevallos}},\ and\ \bibinfo {author} {\bibfnamefont {R.~J.}\ \bibnamefont
  {Cava}},\ }\href {https://doi.org/10.1088/2053-1591/aa60a2} {\bibfield
  {journal} {\bibinfo  {journal} {Materials Research Express}\ }\textbf
  {\bibinfo {volume} {4}},\ \bibinfo {pages} {036102} (\bibinfo {year}
  {2017})}\BibitemShut {NoStop}%
\bibitem [{\citenamefont {Guo}\ \emph {et~al.}(2019)\citenamefont {Guo},
  \citenamefont {Ghasemi}, \citenamefont {Broholm},\ and\ \citenamefont
  {Cava}}]{CavaPRM2019}%
  \BibitemOpen
  \bibfield  {author} {\bibinfo {author} {\bibfnamefont {S.}~\bibnamefont
  {Guo}}, \bibinfo {author} {\bibfnamefont {A.}~\bibnamefont {Ghasemi}},
  \bibinfo {author} {\bibfnamefont {C.~L.}\ \bibnamefont {Broholm}},\ and\
  \bibinfo {author} {\bibfnamefont {R.~J.}\ \bibnamefont {Cava}},\ }\href
  {https://doi.org/10.1103/PhysRevMaterials.3.094404} {\bibfield  {journal}
  {\bibinfo  {journal} {Phys. Rev. Materials}\ }\textbf {\bibinfo {volume}
  {3}},\ \bibinfo {pages} {094404} (\bibinfo {year} {2019})}\BibitemShut
  {NoStop}%
\bibitem [{\citenamefont {Moon}\ \emph {et~al.}(1968)\citenamefont {Moon},
  \citenamefont {Koehler}, \citenamefont {Child},\ and\ \citenamefont
  {Raubenheimer}}]{MoonPR1968}%
  \BibitemOpen
  \bibfield  {author} {\bibinfo {author} {\bibfnamefont {R.~M.}\ \bibnamefont
  {Moon}}, \bibinfo {author} {\bibfnamefont {W.~C.}\ \bibnamefont {Koehler}},
  \bibinfo {author} {\bibfnamefont {H.~R.}\ \bibnamefont {Child}},\ and\
  \bibinfo {author} {\bibfnamefont {L.~I.}\ \bibnamefont {Raubenheimer}},\
  }\href@noop {} {\bibfield  {journal} {\bibinfo  {journal} {Phys. Rev.}\
  }\textbf {\bibinfo {volume} {176}} (\bibinfo {year} {1968})}\BibitemShut
  {NoStop}%
\bibitem [{\citenamefont {Mu}\ \emph {et~al.}(2007{\natexlab{a}})\citenamefont
  {Mu}, \citenamefont {Wang}, \citenamefont {Shan},\ and\ \citenamefont
  {Wen}}]{GangPRB2007}%
  \BibitemOpen
  \bibfield  {author} {\bibinfo {author} {\bibfnamefont {G.}~\bibnamefont
  {Mu}}, \bibinfo {author} {\bibfnamefont {Y.}~\bibnamefont {Wang}}, \bibinfo
  {author} {\bibfnamefont {L.}~\bibnamefont {Shan}},\ and\ \bibinfo {author}
  {\bibfnamefont {H.-H.}\ \bibnamefont {Wen}},\ }\href
  {https://doi.org/10.1103/PhysRevB.76.064527} {\bibfield  {journal} {\bibinfo
  {journal} {Phys. Rev. B}\ }\textbf {\bibinfo {volume} {76}},\ \bibinfo
  {pages} {064527} (\bibinfo {year} {2007}{\natexlab{a}})}\BibitemShut
  {NoStop}%
\bibitem [{\citenamefont {Wang}\ \emph {et~al.}(2007)\citenamefont {Wang},
  \citenamefont {Yan}, \citenamefont {Shan}, \citenamefont {Wen}, \citenamefont
  {Tanabe}, \citenamefont {Adachi},\ and\ \citenamefont {Koike}}]{WangPRB2007}%
  \BibitemOpen
  \bibfield  {author} {\bibinfo {author} {\bibfnamefont {Y.}~\bibnamefont
  {Wang}}, \bibinfo {author} {\bibfnamefont {J.}~\bibnamefont {Yan}}, \bibinfo
  {author} {\bibfnamefont {L.}~\bibnamefont {Shan}}, \bibinfo {author}
  {\bibfnamefont {H.-H.}\ \bibnamefont {Wen}}, \bibinfo {author} {\bibfnamefont
  {Y.}~\bibnamefont {Tanabe}}, \bibinfo {author} {\bibfnamefont
  {T.}~\bibnamefont {Adachi}},\ and\ \bibinfo {author} {\bibfnamefont
  {Y.}~\bibnamefont {Koike}},\ }\href
  {https://doi.org/10.1103/PhysRevB.76.064512} {\bibfield  {journal} {\bibinfo
  {journal} {Phys. Rev. B}\ }\textbf {\bibinfo {volume} {76}},\ \bibinfo
  {pages} {064512} (\bibinfo {year} {2007})}\BibitemShut {NoStop}%
\bibitem [{\citenamefont {Mu}\ \emph {et~al.}(2007{\natexlab{b}})\citenamefont
  {Mu}, \citenamefont {Wang}, \citenamefont {Shan},\ and\ \citenamefont
  {Wen}}]{MuPRB2007}%
  \BibitemOpen
  \bibfield  {author} {\bibinfo {author} {\bibfnamefont {G.}~\bibnamefont
  {Mu}}, \bibinfo {author} {\bibfnamefont {Y.}~\bibnamefont {Wang}}, \bibinfo
  {author} {\bibfnamefont {L.}~\bibnamefont {Shan}},\ and\ \bibinfo {author}
  {\bibfnamefont {H.-H.}\ \bibnamefont {Wen}},\ }\href
  {https://doi.org/10.1103/PhysRevB.76.064527} {\bibfield  {journal} {\bibinfo
  {journal} {Phys. Rev. B}\ }\textbf {\bibinfo {volume} {76}},\ \bibinfo
  {pages} {064527} (\bibinfo {year} {2007}{\natexlab{b}})}\BibitemShut
  {NoStop}%
\bibitem [{\citenamefont {Adhikari}\ \emph {et~al.}(2019)\citenamefont
  {Adhikari}, \citenamefont {Shen}, \citenamefont {Kunwar}, \citenamefont
  {Jeon}, \citenamefont {Maple}, \citenamefont {Dzero},\ and\ \citenamefont
  {Almasan}}]{AdhikariPRB2019}%
  \BibitemOpen
  \bibfield  {author} {\bibinfo {author} {\bibfnamefont {R.~B.}\ \bibnamefont
  {Adhikari}}, \bibinfo {author} {\bibfnamefont {P.}~\bibnamefont {Shen}},
  \bibinfo {author} {\bibfnamefont {D.~L.}\ \bibnamefont {Kunwar}}, \bibinfo
  {author} {\bibfnamefont {I.}~\bibnamefont {Jeon}}, \bibinfo {author}
  {\bibfnamefont {M.~B.}\ \bibnamefont {Maple}}, \bibinfo {author}
  {\bibfnamefont {M.}~\bibnamefont {Dzero}},\ and\ \bibinfo {author}
  {\bibfnamefont {C.~C.}\ \bibnamefont {Almasan}},\ }\href
  {https://doi.org/10.1103/PhysRevB.100.174509} {\bibfield  {journal} {\bibinfo
   {journal} {Phys. Rev. B}\ }\textbf {\bibinfo {volume} {100}},\ \bibinfo
  {pages} {174509} (\bibinfo {year} {2019})}\BibitemShut {NoStop}%
\bibitem [{\citenamefont {Rau}\ \emph {et~al.}(2016)\citenamefont {Rau},
  \citenamefont {Wu}, \citenamefont {May}, \citenamefont {Poudel},
  \citenamefont {Winn}, \citenamefont {Garlea}, \citenamefont {Huq},
  \citenamefont {Whitfield}, \citenamefont {Taylor}, \citenamefont {Lumsden},
  \citenamefont {Gingras},\ and\ \citenamefont {Christianson}}]{RauPRL2016}%
  \BibitemOpen
  \bibfield  {author} {\bibinfo {author} {\bibfnamefont {J.~G.}\ \bibnamefont
  {Rau}}, \bibinfo {author} {\bibfnamefont {L.~S.}\ \bibnamefont {Wu}},
  \bibinfo {author} {\bibfnamefont {A.~F.}\ \bibnamefont {May}}, \bibinfo
  {author} {\bibfnamefont {L.}~\bibnamefont {Poudel}}, \bibinfo {author}
  {\bibfnamefont {B.}~\bibnamefont {Winn}}, \bibinfo {author} {\bibfnamefont
  {V.~O.}\ \bibnamefont {Garlea}}, \bibinfo {author} {\bibfnamefont
  {A.}~\bibnamefont {Huq}}, \bibinfo {author} {\bibfnamefont {P.}~\bibnamefont
  {Whitfield}}, \bibinfo {author} {\bibfnamefont {A.~E.}\ \bibnamefont
  {Taylor}}, \bibinfo {author} {\bibfnamefont {M.~D.}\ \bibnamefont {Lumsden}},
  \bibinfo {author} {\bibfnamefont {M.~J.~P.}\ \bibnamefont {Gingras}},\ and\
  \bibinfo {author} {\bibfnamefont {A.~D.}\ \bibnamefont {Christianson}},\
  }\href {https://doi.org/10.1103/PhysRevLett.116.257204} {\bibfield  {journal}
  {\bibinfo  {journal} {Phys. Rev. Lett.}\ }\textbf {\bibinfo {volume} {116}},\
  \bibinfo {pages} {257204} (\bibinfo {year} {2016})}\BibitemShut {NoStop}%
\bibitem [{\citenamefont {Xie}\ \emph {et~al.}(2012)\citenamefont {Xie},
  \citenamefont {Su},\ and\ \citenamefont {Li}}]{XiePhysica2012}%
  \BibitemOpen
  \bibfield  {author} {\bibinfo {author} {\bibfnamefont {L.}~\bibnamefont
  {Xie}}, \bibinfo {author} {\bibfnamefont {T.}~\bibnamefont {Su}},\ and\
  \bibinfo {author} {\bibfnamefont {X.}~\bibnamefont {Li}},\ }\href
  {https://doi.org/https://doi.org/10.1016/j.physc.2012.04.037} {\bibfield
  {journal} {\bibinfo  {journal} {Physica C: Superconductivity}\ }\textbf
  {\bibinfo {volume} {480}},\ \bibinfo {pages} {14 } (\bibinfo {year}
  {2012})}\BibitemShut {NoStop}%
\bibitem [{\citenamefont {Mahdavifar}\ and\ \citenamefont
  {Akbari}(2008)}]{AkbariJPCM2008}%
  \BibitemOpen
  \bibfield  {author} {\bibinfo {author} {\bibfnamefont {S.}~\bibnamefont
  {Mahdavifar}}\ and\ \bibinfo {author} {\bibfnamefont {A.}~\bibnamefont
  {Akbari}},\ }\href {https://doi.org/10.1088/0953-8984/20/21/215213}
  {\bibfield  {journal} {\bibinfo  {journal} {Journal of Physics: Condensed
  Matter}\ }\textbf {\bibinfo {volume} {20}},\ \bibinfo {pages} {215213}
  (\bibinfo {year} {2008})}\BibitemShut {NoStop}%
\bibitem [{\citenamefont {Politi}\ \emph {et~al.}(2006)\citenamefont {Politi},
  \citenamefont {Pini},\ and\ \citenamefont {Stamps}}]{PolitiPRB2006}%
  \BibitemOpen
  \bibfield  {author} {\bibinfo {author} {\bibfnamefont {P.}~\bibnamefont
  {Politi}}, \bibinfo {author} {\bibfnamefont {M.~G.}\ \bibnamefont {Pini}},\
  and\ \bibinfo {author} {\bibfnamefont {R.~L.}\ \bibnamefont {Stamps}},\
  }\href {https://doi.org/10.1103/PhysRevB.73.020405} {\bibfield  {journal}
  {\bibinfo  {journal} {Phys. Rev. B}\ }\textbf {\bibinfo {volume} {73}},\
  \bibinfo {pages} {020405} (\bibinfo {year} {2006})}\BibitemShut {NoStop}%
\end{thebibliography}%

\end{document}